# Planar-type silicon thermoelectric generator with phononic nanostructures for 100 μW energy harvesting

Ryoto Yanagisawa[1]*, Sota Koike[1], Tomoki Nawae[1], Naohito Tsujii[2], Yanan Wang[2,3], Takao Mori[2,3], Patrick Ruther[4], Oliver Paul[4], Yoshifumi Yoshida[5], Junichi Harashima[6], Takashi Kinumura[7], Yuta Inada[7], and Masahiro Nomura[1]*

[1]Institute of Industrial Science, The University of Tokyo, Tokyo 153-8505, Japan

[2]Research Center for Material Nanoarchitectonics (WPI-MANA), National Institute for Material Science, Tsukuba 305-0044, Japan

[3]Graduate School of Pure and Applied Science, University of Tsukuba, Tsukuba 305-8671, Japan

[4]Department of Microsystems Engineering (IMTEK), University of Freiburg, 79110 Freiburg, Germany

[5]Seiko Future Creation Inc., Chiba 270-2222, Japan

[6]Toppan Inc, Tokyo 112-8531, Japan

[7]ICI center, Maeda Corporation, Toride 302-0021, Japan

*Correspondence should be addressed to R.Y. (e-mail: r-yanagi@iis.u-tokyo.ac.jp) or M.N. (email: nomura@iis.u-tokyo.ac.jp)

*Energy harvesting is essential for the internet-of-things networks where a tremendous number of sensors require power. Thermoelectric generators (TEGs), especially those based on silicon (Si), are a promising source of clean and sustainable energy for these sensors. However, the reported performance of planar-type Si TEGs never exceeded power factors of $0.1\ \mu W\,cm^{-2}K^{-2}$ due to the poor thermoelectric performance of Si and the suboptimal design of the devices. Here, we report a planar-type Si TEG with a power factor of $1.3\ \mu W\,cm^{-2}K^{-2}$ around room temperature. The increase in thermoelectric performance of Si by nanostructuring based on the phonon-glass electron-crystal concept and optimized three-dimensional heat-guiding structures resulted in a significant power factor. In-field testing demonstrated that our Si TEG functions as a 100-μW-class harvester. This result is an essential step toward energy harvesting with a low-environmental load and cost-effective material with high throughput, a necessary condition for energy-autonomous sensor nodes for the trillion sensors universe.*

## Introduction

Improvements in computing speed and low power consumption of electronic components have led to the use of numerous sensors around us, enriching human life. In the internet of things society, where smart devices are connected through networks, energy-autonomous sensors are essential elements. Energy harvesting technology uses ambient energy sources such as mechanical vibration, heat, friction, and light[1–3]. These technologies can also contribute to carbon neutrality by leveraging their respective strengths. Thermoelectric (TE) conversion, one of the most promising energy harvesting technologies, generates electricity from thermal gradients[4,5]. Thermoelectric generators (TEGs) are known for their robustness because they are all solid-state devices with no moving parts. Since heat sources exist in various locations, sensor nodes with TE harvesters are practical if their power generation is sufficient.

The figure of merit $ZT$ of TE materials is expressed by the Seebeck coefficient $S$, electrical conductivity $\sigma$, thermal conductivity $\kappa$, and temperature $T$ as $ZT = S^2 \sigma T / \kappa$. High values of $ZT > 1$ have been reported for Bi, Te, Sb, and Pb compounds[6–8]. The compound $Bi_2Te_3$ is widely used in practical applications[9–11]. In some complex compounds of heavy metals, high Seebeck coefficients due to the large effective mass of electrons and low thermal conductivities due to phonon scattering have resulted in high $ZT$ values. For energy harvesting applications, it is further advantageous that the materials are readily available, inexpensive, have a low environmental impact, and can be supplied in large quantities[12,13].

Silicon (Si) is the most widely used material in electronics and meets the above conditions[14,15]. However due to its high thermal conductivity $\kappa > 130 \text{ Wm}^{-1}\text{K}^{-1}$, its $ZT$ value is as low as 0.001 in bulk material at room temperature[16]. Therefore, it has been considered unsuitable as a TE material. However, researchers have reduced the thermal conductivity by using nanostructures[17,18], and achieved $ZT > 0.1$ at room temperature, such as 0.6 in nanowires (NWs)[19,20], 0.4 in porous thin-films[21], 0.3 in and 0.2 in alloys with germanium[22,23]. While improvements in material performance have been reported, there have not been as many studies on planar device structures fabricated on Si wafers, which is necessary to move to a practical stage using mass-producible Si technology.

Planar-type TEGs fabricated on Si wafers can be vertical or hybrid, depending on the orientation of the TE material[24,25]. In the hybrid type, TEGs generate power by extracting the temperature difference $\Delta T_{TE}$ in the in-plane direction of the TE material from the temperature difference $\Delta T$ in the out-of-plane direction of the wafer. The performance of planar-type TEG is evaluated by the area density of power generation per temperature difference (the power factor of the device in a unit of $\text{Wm}^{-2}\text{K}^{-2}$). Applying a $\Delta T_{TE}$ with an efficiency ratio $\eta_{\Delta T} = \Delta T_{TE}/\Delta T$ is essential in forming a suspension bridge structure. $\Delta T_{TE}$ can be increased by making the bridges longer and the material thinner. However, there is a trade-off in that the internal electrical resistance of the device increases, and an optimization of the device structure, both thermal and electrical, is required. In previous reports, there was a limit to the performance improvement by optimizing the dimensions of the poly-Si film, and the power factor of the device has been limited to about 0.1 µWcm$^{-2}$K$^{-2}$ due to the significant electrical resistance[24–28]. There are reports of about 0.25 µWcm$^{-2}$K$^{-2}$ in devices using SiGe instead of poly-Si[29,30], and 0.3 µWcm$^{-2}$K$^{-2}$ in devices using bundles of Si NWs as bridges[31]. However, material cost and compatibility with wafer processes have been an issue.

This study demonstrates a planar-type Si TEG that achieves the power factor of 1.3 µWcm$^{-2}$K$^{-2}$ using poly-Si thin films as the TE material. This is more than ten times higher power density than previously reported achieved by enhancing the thermoelectric performance through the fabrication of phononic nanostructures in the thermoelectric material and a structural design that optimizes electrical and thermal resistance. In particular, the three-dimensional structure for directing the heat flow from the cross-plane to the in-plane direction enables $\eta_{\Delta T} \geq 30\%$, contributing significantly to performance improvement. We then demonstrate that our Si TEG produces an average power of 100 µW in an in-field measurement over four days.

## Planar-type nano phononic Si thermoelectric generator

In a planar-type TEG, the TE material parts that generate electricity have a thin film shape in the in-plane direction, with a double-cavity structure above and below the thin film (Fig. 1a,b). This structure allows the thermoelectric thin film to be suspended in mid-air to achieve a large temperature difference in the in-plane direction, thereby increasing the power generation density. Figure 1c shows a scanning electron microscope (SEM) image of a unit cell of TEG with an electric circuit when a temperature difference $\Delta T_{TE}$ is induced in the thin film. The performance of the TEG is evaluated by the power density $P$, expressed by the TE voltage $V_{TE}$, electrical resistance $R$, and device area $A$ as $P = V_{TE}^2/(4RA)$. Ideally, the power density does not depend on the number of devices integrated. However, the design of the TEG should not be based on the device alone but rather on the configuration of the number of devices in series and in parallels, which are optimized by considering the circuit design of the energy harvesting module. Therefore, based on an optimization calculations for 2400 units, we fabricated the device as a parallel arrangement of 20 lines, each with 120 units in series (Fig. 1d). The details of the fabrication and structure of the TEG are given in the Methods section and Supplementary Information 1.

In this study, two essential factors enabled the enormous increase achieved in power density: the enhancement of $ZT$ by nanostructuring and the three-dimensional heat flow control of the device. Poly-Si thin film thermoelectric materials have nanostructures based on phonon engineering for $ZT$ enhancement. The phononic crystal (PnC) nanostructure is designed considering the mean free path of charge and heat carriers, known as the concept of PGEC (phonon-glass electron-crystal) proposed by Slack in 1995[32]. In this structure, the phonon transport is significantly reduced while the electron transport is minimally disturbed. In porous structures, such as PnCs, the distance between the sidewalls of the pores (neck size $n$) is an essential structural parameter (Fig. 1e). The film thickness is also a critical structural parameter in TEG. If the film is too thick, the temperature difference will not be achieved, resulting in low power density, while nanofabrication will be difficult. On the other hand, if it is too thin, the electrical resistance becomes too high, and furthermore the mechanical strength, which is essential for practical use, cannot be ensured. In this study, we have formed PnC nanostructures with circular through-holes in a 1.1-μm-thick poly-Si film with a periodicity of 300 nm in a honeycomb lattice. The neck sizes range from 8 nm to 116 nm (e.g., Fig. 1f shows a PnC with $n$ = 8 nm). The dependence of thermal conductivity, electrical conductivity, and power generation on mechanical strength was investigated.

The other important factor, the heat flow control in the device, is the optimization of the structural parameters in the thin film and the optimization of the structure perpendicular to the thin film. The former was optimized using the finite element method with the thermal and electrical parameters of the material and boundary resistances obtained by our measurements or database[33] (Supplementary Information 2). Here, we describe a double-cavity structure that increases the power density by applying as large a temperature difference as possible to the thermoelectric material. In addition, we have developed a fabrication technique for heat-guiding structures by cap wafer bonding. Creating a heat flow path from the hot to the cold side through solid-solid interfaces is essential.

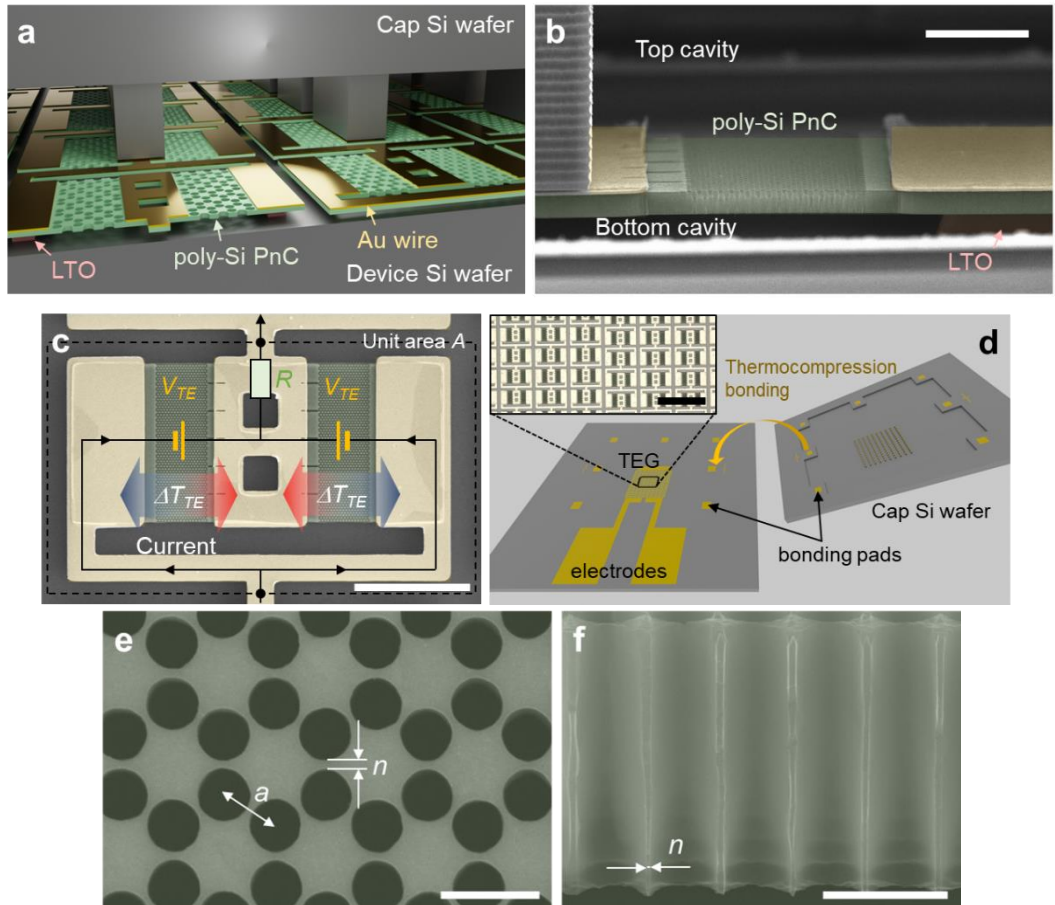

**Fig. 1 | Planar-type nano phononic Si thermoelectric generator. a**, Schematic 3D picture of TEGs composed of a device and cap wafer. **b**, Side-view SEM images of TEG after cap wafer bonding. Scale bar, 5 μm. **c**, SEM image of a unit cell of TEG. Scale bar, 20 μm. **d**, Schematic 3D picture of fabricated device and cap wafers before bonding. The inset shows an optical microscope image of connected TEG cells. Scale bar, 100 μm. **e,f**, SEM images of PnC nanostructures from the top and side view. Scale bars, 500 nm.

## Nanostructured polycrystalline Si TE material

Although various structural parameters can be approximately optimized by the finite element method, the dynamics of charge and heat carriers in nanostructures below 100 nm and systems with strong surface scattering are not straightforward. Therefore, it is necessary to experimentally investigate the structures that maximize the power density, starting from those found in simulations under conditions where mechanical stability is ensured. We have performed thermal and electrical conductivity measurements on several n-type nanostructured poly-Si films with different neck sizes to evaluate $ZT$ (Supplementary Information 3). We first measured the carrier concentration, Seebeck coefficient, and electrical conductivity of the poly-Si film before nanostructuring. We found that the carrier concentration was $2.2 \times 10^{20}$ cm$^{-3}$, the Seebeck coefficient was $-100$ μVK$^{-1}$, the electrical conductivity was 166 kSm$^{-1}$ and the material power factor was 1.66 mWm$^{-1}$K$^{-2}$. Figure 2a shows the measured thermal conductivity as a function of the neck size. The dashed line shows the thermal conductivity of the film with no nanostructure (31 Wm$^{-1}$K$^{-1}$). It is reduced to less than a quarter of that of single-crystal bulk Si due to the surface scattering in a thin film and polycrystalline grain boundary scattering. As the neck size decreases, the thermal conductivity decreases, with a thermal conductivity of 9 Wm$^{-1}$K$^{-1}$ for a sample with a neck width of 36 nm. We fabricated structures with a neck size $n$ of about 8 nm or less and obtained further reduced thermal conductivities, but the mechanical strength was insufficient for device fabrication. Thermal phonons are the main heat carriers in this sample, and phonon scattering is the sum of boundary scattering, Umklapp scattering, and impurity scattering as described by the Matthiessen rule. At around room temperature, the Umklapp scattering has a larger influence on the bulk material, while the boundary scattering becomes larger and dominant in the nanostructures.

Figure 2b shows the electrical conductivity versus the neck width of the PnC nanostructures. The electrical conductivity is also reduced, more strongly than expected, by fabricating the circular holes, especially in the neck size below 60 nm. This reduction could be due to surface oxidation, which may have reduced the effective neck size. In line with the PGEC concept, the fabrication of PnC nanostructures, although smaller than expected, results in a larger $\sigma/\kappa$ ratio. However, there is room to further increase the $ZT$ by analyzing and improving the surface conditions. A plot of the $ZT$ versus neck width is shown in Fig. 2c. We assumed that the Seebeck coefficient is constant irrespective of the nanostructure size[20,21]. Nanostructured poly-Si shows a $ZT$ of $2.0 \times 10^{-2}$ at 300 K, which is 20 times better than bulk Si. Figure 2d shows the grain size distribution estimated by the transmission electron microscope (TEM) observation. The grain size of poly-Si is dominated by sizes around 50 - 100 nm, and some grain boundaries are in the necks causing the reduction of $\sigma$ too. Making the smaller PnCs in the poly-Si with larger grain size might improve $\sigma/\kappa$ more efficiently[34,35] (Supplementary Information 4).

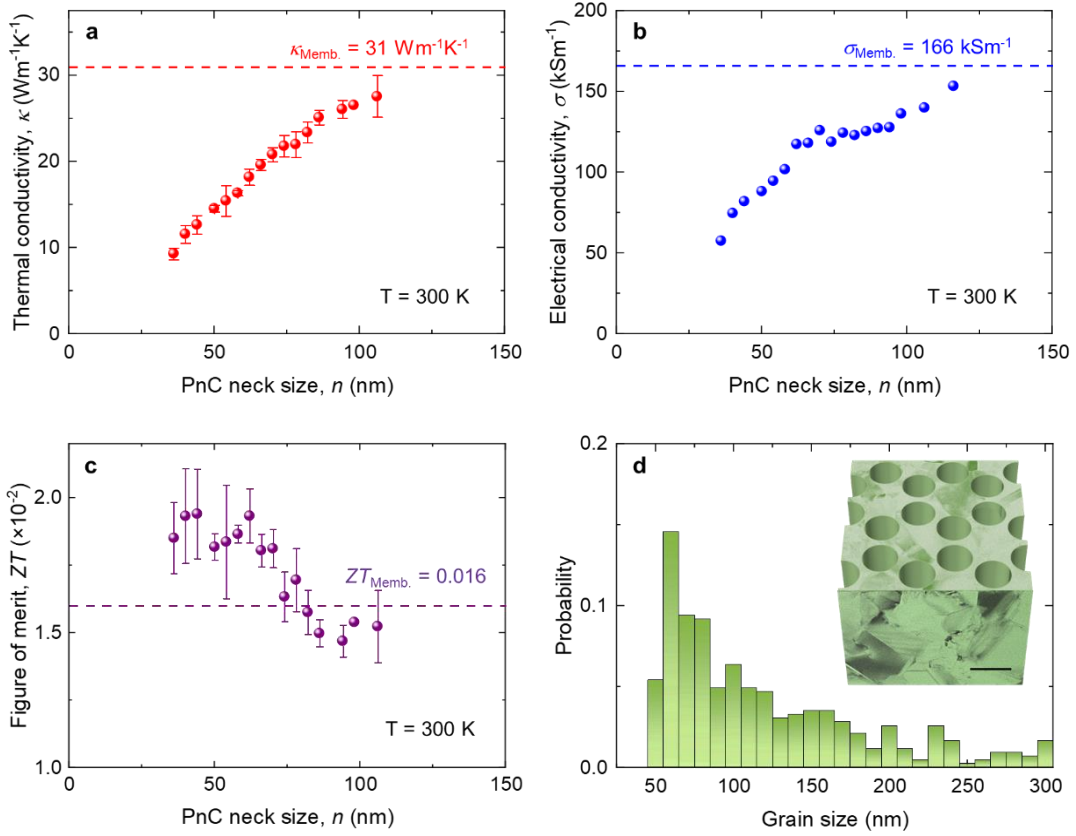

**Fig. 2 | Nano phononic Si thermoelectric material. a-c**, Measured thermoelectric properties of the nanostructured poly-Si membrane as a function of PnC neck size: thermal conductivity (**a**), electrical conductivity (**b**), and thermoelectric figure of merit (**c**). **d**, Histogram of the grain size measured by surface and cross-sectional TEM observations. The inset shows two TEM images overlaid with a PnC pattern. Scale bar, 300 nm.

## Characterization of planar-type Si TEG

We evaluated the performance of the fabricated 120 series × 20 parallel TEG at room temperature. The details of the measurement setup are described in Methods and Supplementary Information 5. The area of the 2400 units is 0.077 cm$^2$, including the wiring between the units. The internal electrical resistance of the TEG was 38.8 Ω, designed to be less than 50 Ω for a boost circuit in the latter stage of the energy harvesting module. Next, the TEG voltage was measured when an external temperature difference $\Delta T$ was applied: an open circuit voltage of 3.9 mV was measured at $\Delta T = 1$ K, yielding a voltage factor of 50.6 mVcm$^{-2}$K$^{-1}$ (Fig. 3a). The TEG voltage is proportional to the applied temperature difference in the range of about 15 K near room temperature. This fact indicates that the temperature dependence of the physical properties is negligible in this range. The estimated ratio $\eta_{\Delta T} = \Delta T_{TE}/\Delta T$ is as high as 32.4 %, a 160-fold improvement from 0.2 % in our previous TEG without a cap wafer[36]. While the total heat flow path length is 1 mm, consisting mainly of the Si substrate and a cap wafer (525 μm-thick, respectively), the PnC poly-Si with a length of only 9 μm concentrates a temperature difference of up to 32 %. The fact that the amount of power generated is proportional to the square of $\varDelta T_{TE}$ shows that our double-cavity structure is highly effective in increasing the power density.

From the measured TEG voltage and the internal electrical resistance, we evaluated the power density $P$, which is proportional to the square of the temperature difference $\varDelta T$, and obtained a power factor of the TEG of 1.3 $\mu Wcm^{-2}K^{-2}$ (Fig. 3b). To our knowledge, this is the highest performance for a hybrid planar-type Si thermoelectric device. A power generation density of 100 $\mu Wcm^{-2}$ can be obtained from a temperature difference of less than 10 K, which is common in the everyday environment. In the inset of Figure 3b, we show the IR camera image to characterize the $\varDelta T$ in the environment. We fixed the test case body with planar-type TEG and Al heat-sink on the backside of the solar cell panel and observed $\varDelta T = 11.8$ K while the panel was irradiated by the sun. Although some $\varDelta T$ is lost at the interfaces between the TEG and the heat-sink, we expect the TE energy harvester to be applicable.

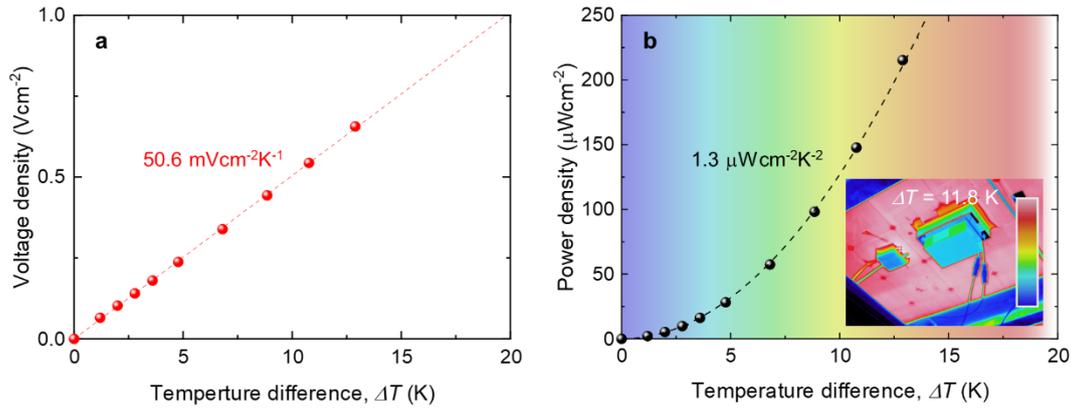

**Fig. 3 | 100 μW Si thermoelectric energy harvester. a,b**, Output voltage (**a**) and power density (**b**) of TEG as a function of applied temperature difference $\varDelta T$. The inset shows an IR camera image of the temperature map around the TEG module. Color bar, from 20 to 40 °C.

## Benchmark of planar-type Si TEG for energy harvesting

To clarify the characteristics and performance of our device, we will discuss it in comparison with previously reported devices in the past two decades in terms of device power factor. Due to the practical superiority of Si, TEGs based on poly-Si films have been studied to improve the power factor and to develop fabrication methods. Recently, a power factor of 0.1 $\mu Wcm^{-2}K^{-2}$ has been reported for devices fabricated using only processes compatible with the BiCMOS process[28]. There have also been reports of devices based on single-crystalline Si NWs, with a power factor of 0.5 $\mu Wcm^{-2}K^{-2}$ using short NWs of less than 1 μm in length on the $SiO_2$ layer[37]. To increase the power density by utilizing large $\varDelta T$, making a bridge structure with bottom and top cavities is essential. Using bundles of Si or SiGe NWs, devices have been reported with a power factor of about 0.3 $\mu Wcm^{-2}K^{-2}$ [31]. Devices using poly-SiGe films, which have a higher bulk performance than Si, have also been reported at 0.25 µWcm$^{-2}$K$^{-2}$ due to their low thermal conductivity. However, the high material cost of germanium and the low performance of n-type films have remained challenging.

Generally, both n- and p-type materials are used in TEGs. In contrast, our TEG is a uni-leg type, so the fabrication process is simplified and less costly. In addition, n-type material has relatively higher electrical conductivity and lower contact resistance to metals than p-type material, which reduces electrical resistance for better circuit compatibility. Furthermore, optimization of the through-hole with a small neck size and three-dimensional heat-guiding structures enabled a power factor of 1.3 µWcm$^{-2}$K$^{-2}$, which is more than ten times higher than similar poly-Si films.

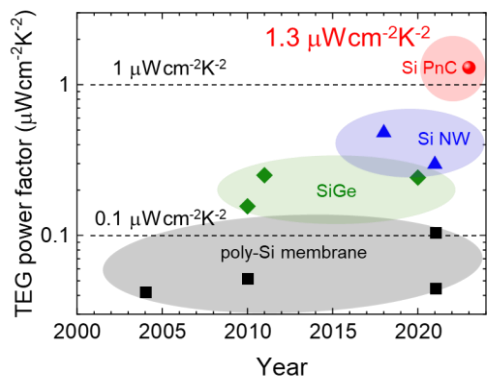

**Fig. 4 | Benchmark of planar-type Si-based TEG.** Comparison of TEG power factors among hybrid planar-type Si devices reported in the last decades. Colors represent the different material groups. Black, poly-Si membrane[28,38–40]; green, SiGe[29,30,41]; blue, Si NW[31,37]; red Si PnC (this work).

## In-field characterization of energy harvest module

Considering the application of TEG to wireless sensing systems, a simple sensor with low data volume and low-power communication consumes about 5 J for one sensing and data transmission in the case of Bluetooth Low Energy (BLE), which can communicate over several hundred meters. We performed in-field measurements to demonstrate power generation over a long period by an actual $\Delta T$ in the environment. The solar cell panel was used for the power generation measurement. As shown in Figs. 5a and 5b, two thermocouples and a Si TEG module were installed on the backside of the solar panel. The two thermocouples measure the temperature difference $\Delta T$ between the surface of the panel and the atmosphere 5 cm away from the panel, as shown in Fig. 5c. The output voltage of the TEG was also measured. Careful thermal design of the module is essential for an efficient thermoelectric power generation. A plastic resin ($\kappa \sim 0.18$ Wm$^{-1}$K$^{-1}$) was used for the case body for better thermal isolation, and two Al blocks and an Al heat sink were used for efficient heat dissipation on the cold side.[42] The transition of $\Delta T$ was recorded over four consecutive days, as shown in Fig. 5d, and the TEG voltage, which is proportional to $\Delta T$ was also recorded. Figure 5e shows the calculated accumulated energy assuming the device power factor of 1.3 µWcm$^{-2}$K$^{-2}$ and an enlarged area of 10 cm$^2$. The details of the calculation are given in the Methods section. As expected, the amount of generated power depends on the weather. Interestingly, a negative value of $\Delta T$ is observed by radiative cooling. Therefore, we have a chance to harvest more energy by installing an inverter if the energy loss is less than the harvested power at night. The average generated power is 98 µW (721 µW at the peak in a sunny day), and the total energy generated is about 34 J within four days. A single measurement and data transmission via BLE requires about 5 J and 70 J for small data (temperature, humidity, etc.) and large data (10 KB image), respectively. Therefore, data can be collected about twice a day for the former application and 14 times per month for the latter. This in-field measurement shows that Si TEGs have reached a level where they can be used as a power source for self-powered sensing systems.

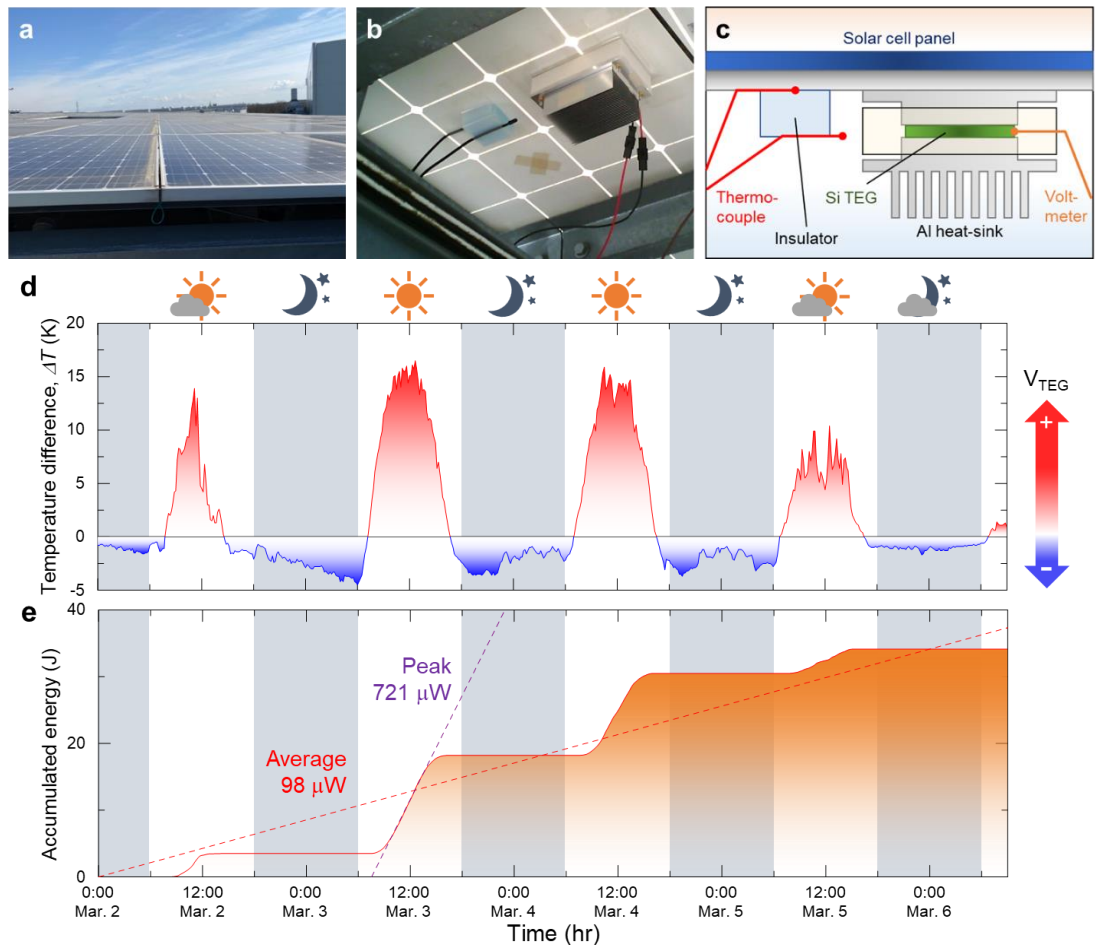

**Fig. 5 | In-field measurement of the environment temperature difference and harvested energy estimation. a,b,** Solar cell panels used in the experiment on the front side (**a**) and the back side with the measurement systems for environmental temperature and TEG voltage (**b**). **c**, Schematic of the measurement system. **d,e,** Measured temperature difference between the backside surface of solar cell panel and air (**d**) and estimated accumulated energy (**e**) over four continuous days in Ibaraki, Japan (35°54'40"N, 140°3'15"E, from 2 to 6 March 2023).

## Conclusions

We demonstrate a planar-type poly-Si TEG with a power factor of 1.3 $\mu Wcm^{-2}K^{-2}$. According to the PGEC concept, the material performance was enhanced by forming through-holes with neck sizes of 8 to 100 nm in a 1.1-μm-thick poly-Si film. In addition, a three-dimensional heat-guiding structure fabricated by bonding the Si cap wafer increases $\eta_{\Delta T}$, and results in high power generation. The device performs ten times better than conventional planar-type Si TEGs and generates more than 200 $\mu Wcm^{-2}$. The four days of in-field power generation demonstrated that the TEG generates an average of close to 100 μW (0.72 mW at the peak) and can be used as an energy-autonomous sensing system even at small temperature differences, such as those obtained from a near-room-temperature environment. The Si device is a uni-leg type that can be mass-produced at a low cost due to its greatly simplified process, which is advantageous for its widespread use. This work can contribute to accelerate the realisation of the internet of things society, which requires an enormous number of sensors.

## Methods

**Planar-type thermoelectric generator fabrication and structure.** The planar-type TEGs were fabricated using lithography-based wafer processes. The n-type 1.1-μm-thick poly-Si layer was deposited by low-pressure chemical vapor deposition at 580 °C with phosphorous as dopant following the deposition of a 1.5-μm-thick low-temperature silicon oxide (LTO) sacrificial layer at 425 °C. The wafer was thermally annealed at 1050 °C for 30 minutes in a nitrogen atmosphere to activate the dopants. The unit device of TEG was designed in a 44 μm × 73 μm area composed of the bridged thermoelectric part and the suspended electrical wire part. The phononic nanostructures were realized in the poly-Si layer using electron beam (EB) lithography and reactive ion etching (RIE) of Si with $SF_6$ and $O_2$ plasma using a spin-coated EB resist (ZEP520A) as masking layer (Supplementary Fig. 1a). The 300-nm-thick Au wires were deposited and formed using laser lithography and EB-assisted physical vapor deposition with 10-nm-thick Cr interfacial layer deposition between poly-Si and Au (Supplementary Fig. 1b). After etching the poly-Si between each unit of TEGs, the LTO sacrificial layer was etched from holes and slits using isotropic etching of silicon oxide with HF and ethanol gases (Supplementary Fig. 1c). The cap wafer was fabricated with Cr and Au depositions for the bonding pads and deep RIE of Si to form 24-μm-deep vertical heat guiding structures (Supplementary Fig. 1d,e). Finally, the cap wafer was bonded on the poly-Si device wafer using a flip-chip bonder. After two wafers were aligned in a horizontal position and contacted through Au pads on both wafers, the device wafer was heated up to 330 °C while applying 30 N pressure. The TEG of 2400 units was fabricated in a 2.64 mm by 2.92 mm area on a Si chip.

**Characterization of the thermoelectric generator.** ADVANCE RIKO Mini-PEM was used to evaluate the power density of the TEG devices (Supplementary Fig. 6). The sample was placed between two copper blocks in a vacuum chamber, and carbon sheets were inserted between the sample and the cupper block on both sides as a thermal interface material. On both sides, a thermocouple was embedded in the copper block and contacted with the carbon sheet in the middle to measure the temperature difference $\Delta T$ between both side of the TEG. On the cold side, the temperature of the copper block was controlled by a water flow cooling system at 25 °C. In addition, the temperature of water at the inlet and outlet and the flow rate was monitored so that the heat flux from the bottom block to the sample could be estimated. On the hot side, the temperature of the top copper block was controlled by a heater and changed from 25 to 55 °C to apply $\Delta T$ values from 0 to 12.9 K in the air.

The sample was electrically connected to a voltage source (ADCMT 6244) and a digital multimeter (KEITHLEY DAQ6510) to measure the open-circuit thermoelectric voltage $V_{TEG}$. Sweeping the applied current at each $\Delta T$, we defined the thermoelectric voltage from the intercept of the $IV$ plot. The power density was calculated as $P = V_{TEG}^2/4R_{TEG}A$, with the measured $V_{TEG}$, the electrical resistance $R_{TEG}$, and occupied area $A$ by the 2400 units.

**Estimation of harvested energy by measured temperature differences.** In order to generate large amounts of energy, we designed a planar-type TEG with a device area of 10 cm². In total, 311,349 TEG units can be integrated with a configuration of 1549 series by 201 parallel connections, achieving 49.8 Ω internal electrical resistance, 50.4 mV/$\Delta T$ TEG voltage coefficient, and 1.27 µWcm$^{-2}$K$^{-2}$ power factor. The number of connections was chosen to keep the electrical resistance below 50 Ω. Based on the thermal design of the module (Supplementary Information 6), we assumed a maximum heat exchange efficiency of 50 % between the measured ambient temperature difference and the $\Delta T$ across the TEG, and the output voltage of the TEG was obtained. The resulting voltage was used as the input to the voltage boost converter, which outputs 4.3 V only for inputs higher than 50 mV, and the output of the 4-day current was obtained from the ratio of the voltages at a load resistance of 50 Ω. The accumulated energy obtained over four days as an energy harvester was estimated from the voltage and current after the voltage booster.

**Acknowledgments**

This work was supported by CREST JST (JPMPCR19Q3), MIRAI JST (JPMJMI19A1), and the Project for Developing Innovation Systems of the MEXT, Japan, Kakenhi (21H04635). A part of this work was supported by NICT Advanced ICT Device Laboratory. A part of this work was supported by Tohoku University Nanofab Platform in MEXT Advanced Research Infrastructure for Materials and Nanotechnology in Japan (JPMXP1222TU0025). We also acknowledge Roman Anufriev for assistance with manuscript preparation and Kentaro Furusawa for assistance with fabrication of nanostructures.

**Author contributions**

R. Y. designed, fabricated, and measured the device and material, analyzed the experimental data, and wrote the article. S. K. designed and made the harvester body and supported in-field measurement. T. N. fabricated and measured the materials. N. T., Y. W., and T. M. measured the material properties. P. R. and O. P. made the material for the device. Y. Y., J. H., T. K., and Y. I. supported design and fabrication of harvester body, and in-field measurement. M. N. contributed to the design of the study, writing, funding acquisition, and supervising the entire work.

Supplementary Information for
Planar-type silicon thermoelectric generator with phononic nanostructures for 100 μW energy harvesting

Ryoto Yanagisawa[1]*, Sota Koike[1], Tomoki Nawae[1], Naohito Tsujii[2], Yanan Wang[2,3], Takao Mori[2,3], Patrick Ruther[4], Oliver Paul[4], Yoshifumi Yoshida[5], Junichi Harashima[6], Takashi Kinumura[7], Yuta Inada[7], and Masahiro Nomura[1]*

[1]Institute of Industrial Science, The University of Tokyo, Tokyo 153-8505, Japan

[2]Research Center for Material Nanoarchitectonics (WPI-MANA), National Institute for Material Science, Tsukuba 305-0044, Japan

[3]Graduate School of Pure and Applied Science, University of Tsukuba, Tsukuba 305-8671, Japan

[4]Department of Microsystems Engineering (IMTEK), University of Freiburg, 79110 Freiburg, Germany

[5]Seiko Future Creation Inc., Chiba 270-2222, Japan

[6]Toppan Inc, Tokyo 112-8531, Japan

[7]ICI center, Maeda Corporation, Toride 302-0021, Japan

*e-mail: r-yanagi@iis.u-tokyo.ac.jp, nomura@iis.u-tokyo.ac.jp

## 1. TEG design and fabrication

We have designed and developed a uni-leg planar-type Si thermoelectric generator (TEG). The device is fabricated on an SOI wafer with an n-type doped polycrystalline Si layer as the active layer. The process involves eight steps: first, doping and annealing are performed simultaneously with SOI growth; then, three lithography steps, two RIE steps, and one metal deposition step for nanostructure fabrication, metal wiring, and inter-unit isolation are performed; and finally, one sacrificial layer etching step is performed (Supplementary Fig. 1a-c). In the case of conventional p-type and n-type bi-leg TEGs, two ion implantation and two annealing processes for p-type and n-type are added to the fabrication process, so the throughput of the fabrication process for uni-leg devices is considered high[1]. Cap wafers, on the other hand, are fabricated on undoped silicon wafers with two lithography steps, one metal deposition, and one RIE of Si (Supplementary Fig. 1d,e). It can be fabricated in fewer steps compared to the method of fabricating the top cavity using sacrificial layer stacking and etching[2].

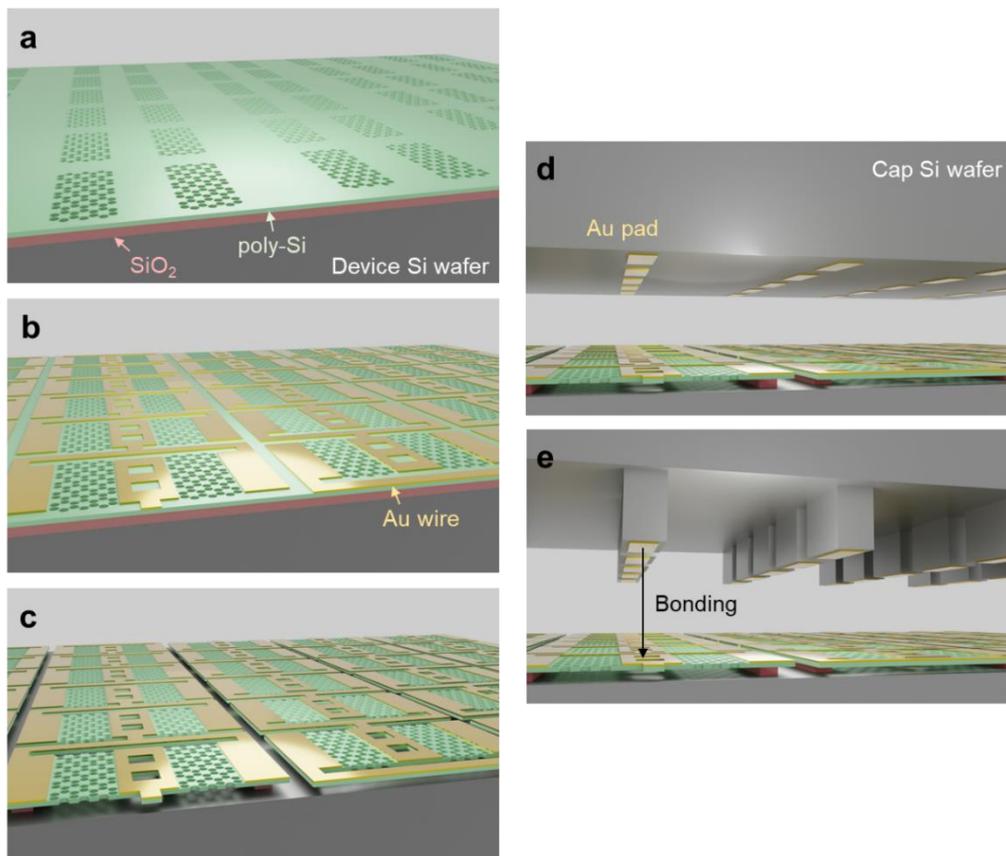

**Supplementary Figure 1 | Fabrication of the planar-type TEG.** Schematic pictures of the TEG in each fabrication step. (**a**) After RIE of PnC nanostructures, (**b**) after PVD of Au metal layers, and (**c**) after RIE of Si for electrical isolation and VHF etching of $SiO_2$ layer to form the bridged structures. (**d**) Fabrication of the cap wafer side after PVD of Au metal pad for bonding step. (**e**) DRIE of Si wafer to form the pillar shape connection for the top cavity structure.

## 2. FEM simulation of TEG power factor

A three-dimensional finite element method (FEM) model was used to simulate the power factor of a planar-type TEG. A steady-state multi-physics solver with heat conduction and current modules was used with the materials listed in Supplementary Tables 1 and 2. As shown in Supplementary Fig. 2, we applied the periodic boundary condition in the X- and Y- directions and a fixed temperature difference $\Delta T$ = 10 K between the top boundary of the cap Si and the bottom of the substrate Si. We also applied a current and measured the electrical resistance of the unit cell. The power density and power factor were calculated from the thermoelectric voltage obtained from the in-plane temperature difference $\Delta T_{TE}$, electrical resistance, and the occupied area of the unit cell. Supplementary Fig. 3 shows the results of simulations with different porosity and bridge lengths of the PnC nanostructures. When the bridge length was changed, the longer the bridge, the more significant the temperature difference $\Delta T_{TE}$ becomes, while the internal electrical resistance and occupied area increase, so there is an optimal bridge length that maximizes the power factor. As the porosity increases, the optimal bridge length shifts to a shorter region, resulting in a larger maximum power factor. When the porosity is 0.44 (corresponding to the case with PnC period = 300 nm, neck size = 44 nm), the power factor reaches a maximum of around 7.6 $\mu Wcm^{-2}K^{-2}$ when the bridge length is about 5 $\mu m$, which is about 30% higher than the performance without nanostructuring. 7.6 $\mu Wcm^{-2}K^{-2}$ power factor of TEG is six times higher than the experimental results, which is explained to be due to the 1.9 times larger in-plane temperature difference $\Delta T_{TE}$ in the simulation and the contribution of additional electrical resistance, such as wiring, in the experiment. The difference in $\Delta T_{TE}$ might come from the thermal resistance between the $\Delta T$ applied by setup and the location where the $\Delta T$ is measured in the experiment, and the $\Delta T$ may have been underestimated in the experiment.

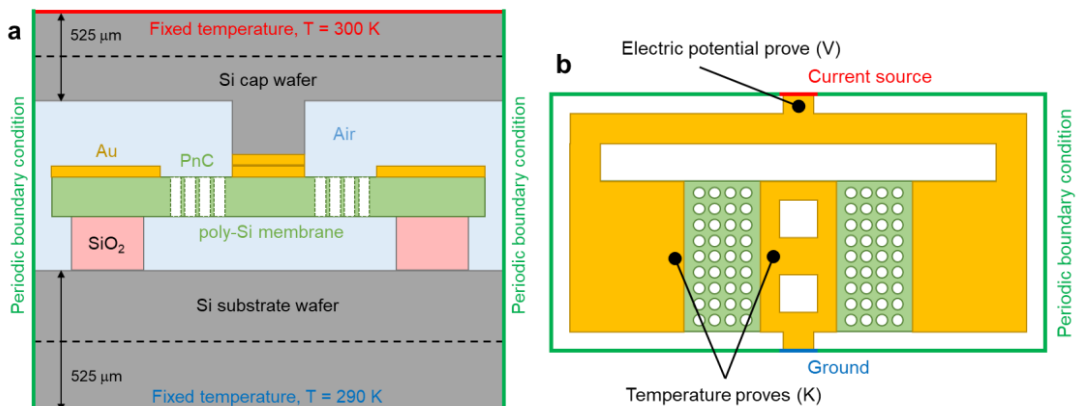

**Supplementary Figure 2 | FEM simulation of planar-type TEG.** (**a**) Side view of a schematic of the FEM model and (**b**) top view of the device layer composed of poly-Si and Au. The model has periodic boundary conditions in the horizontal direction, and a fixed temperature difference is applied to vertical boundaries. The power factor of the device is calculated from simulated thermoelectric voltage and electrical resistance.

**Supplementary Table 1 | Thermoelectric properties of poly-Si PnC used in FEM simulation.**

| Porosity | $\kappa$ (Wm$^{-1}$K$^{-1}$) | $\sigma$ (kSm$^{-1}$) | S ($\mu$VK$^{-1}$) |
|---|---|---|---|
| 0 | 31 | 166 | |
| 0.38 | 18.2 | 117 | 100 |
| 0.44 | 12.7 | 82 | |

**Supplementary Table 2 | Material properties used in FEM simulation.**

| Material | $\kappa$ (Wm$^{-1}$K$^{-1}$) | $\sigma$ (kSm$^{-1}$) | $\rho$ (kgm$^{-3}$) | $C_P$ (Jkg$^{-1}$K$^{-1}$) | $R_{contact}$ ($\Omega\mu m^2$) |
|---|---|---|---|---|---|
| Si | 130 | - | 2329 | 700 | - |
| SiO$_2$ | 1.38 | - | 2203 | 703 | - |
| Au | 317 | 4560 | 19300 | 129 | 100 |
| Air | 0.03 | - | $1.16 \times 10^{-5}$ | 1006 | - |

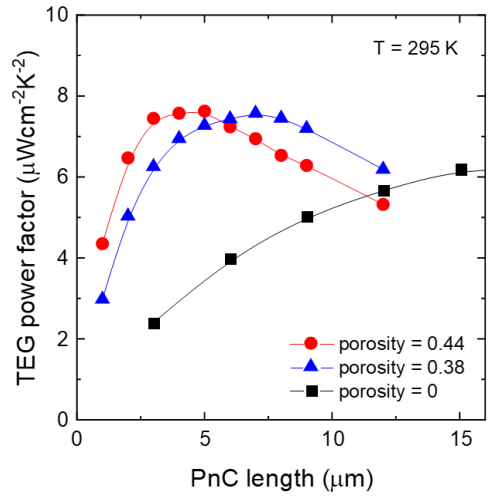

**Supplementary Figure 3 | Simulated TEG power factor.** The plots show the power factor of Si TEG with different PnC sizes as a function of PnC bridge length. As the porosity increase, the optimal PnC length shifts to a shorter range. By making a PnC nanostructure, the maximum TEG power factor can increase.

## 3. Measurement of κ and σ

To measure the thermal and electrical conductivity (κ and σ) of nanostructured poly-Si thin films, we fabricated a sample as shown in Supplementary Fig. 4. The sample consists of bridged nanostructured thin films with a metal pad used in the thermoreflectance method in the center and four electrical measurement terminals connected at both ends. This structure allows κ and σ to be measured for the same nanostructured sample. The κ was measured using the micro-thermoreflectance method, which effectively measures the in-plane κ of thin-film samples. Pump and probe lasers are focused on a metal pad transducer in a vacuum, and the temperature rise due to absorption of the pump laser and the time variation of heat dissipation due to heat conduction through the thin film is measured. From the measured time constant of heat dissipation, κ is determined by FEM simulation using κ as the free fitting parameter[3]. The electrical conductivity σ was measured by the four-probes method. The σ of the nanostructured thin film was determined by FEM simulation using the σ as a free-fitting parameter[4]. Measurements were performed on samples with different nanostructure dimensions, as shown in Supplementary Fig. 4c,d, and consistent data were obtained for samples with nanostructure neck widths ranging from 100 nm to 20 nm.

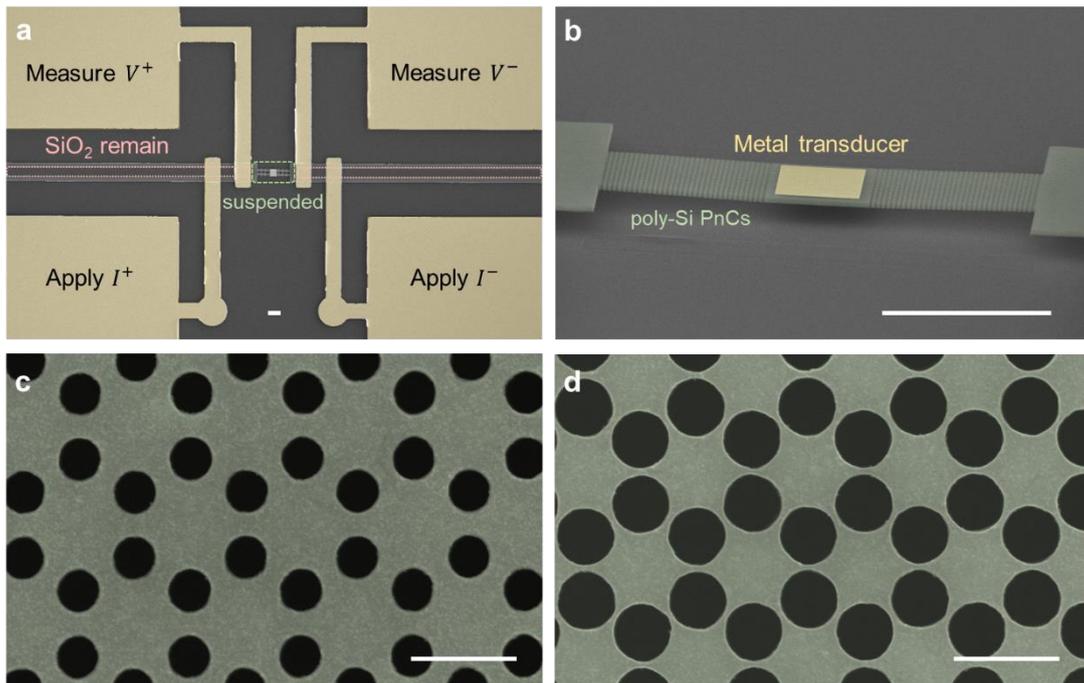

**Supplementary Figure 4 | SEM images of a test structure for κ and σ of poly-Si PnCs.** (**a**) Top comprehensive view and (**b**) tilted close view SEM images of poly-Si PnC test samples for the measurement. Scale bars are 10 μm. SEM images of PnC pattern with (**c**) wide and (**d**) narrow neck size. Scale bars are 500 nm.

## 4. Polycrystalline grain size

A transmission electron microscope with electron diffraction mapping (TEM-EDM) was used to evaluate the grain size of the polycrystalline silicon thermoelectric layer. As shown in Supplementary Fig. 5a,b, the grains are distributed in various sizes in the in-plane and trans-plane directions in a silicon layer of 1.1 μm thickness. The grain size was evaluated from the diameter when the grains were approximated as circles of the same area in the EDM map. The histogram in Fig. 2d shows the smallest grains ranging from about 50 nm to the largest ones over 300 nm, with the size of about 60 nm being the most abundantly distributed. The grain boundaries may overlap with the neck of the PnC nanostructured thin film because it has a cross-section with a thickness of about 1.1 um and a width of 40~120 nm. When a grain boundary exists inside the neck, the thermal and electrical conductivities are expected to be reduced due to boundary scattering of phonons and electrons. This results in a reduction of electrical conductivity larger than expected when the neck width is reduced, so the improvement in *ZT* by the PnC nanostructure is only about 25%. Therefore, appropriate control of the grain size of the silicon layer is expected to recover the electrical conductivity at the neck of the PnC nanostructure and further improve the *ZT*.

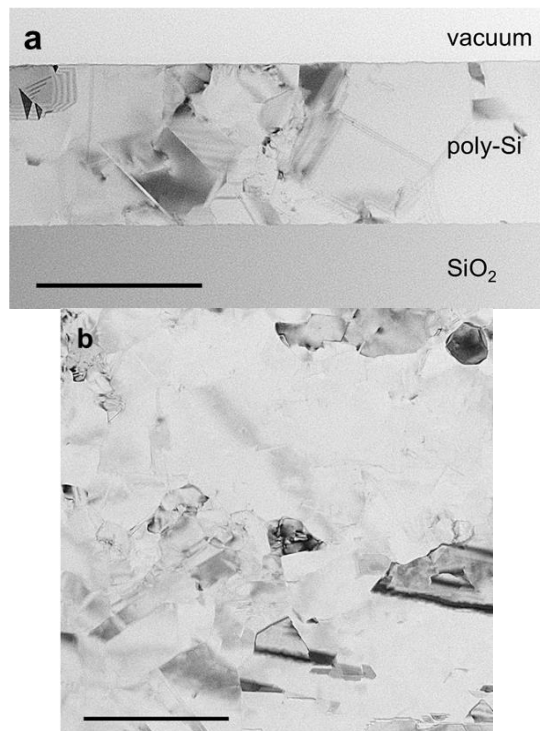

**Supplementary Figure 5 | TEM evaluation of polycrystalline grain sizes.** (**a**) Side view and (**b**) top view TEM images of poly-Si thin-film. Scale bars are 1 μm.

## 5. Picture and schematic of TEG measurement setup

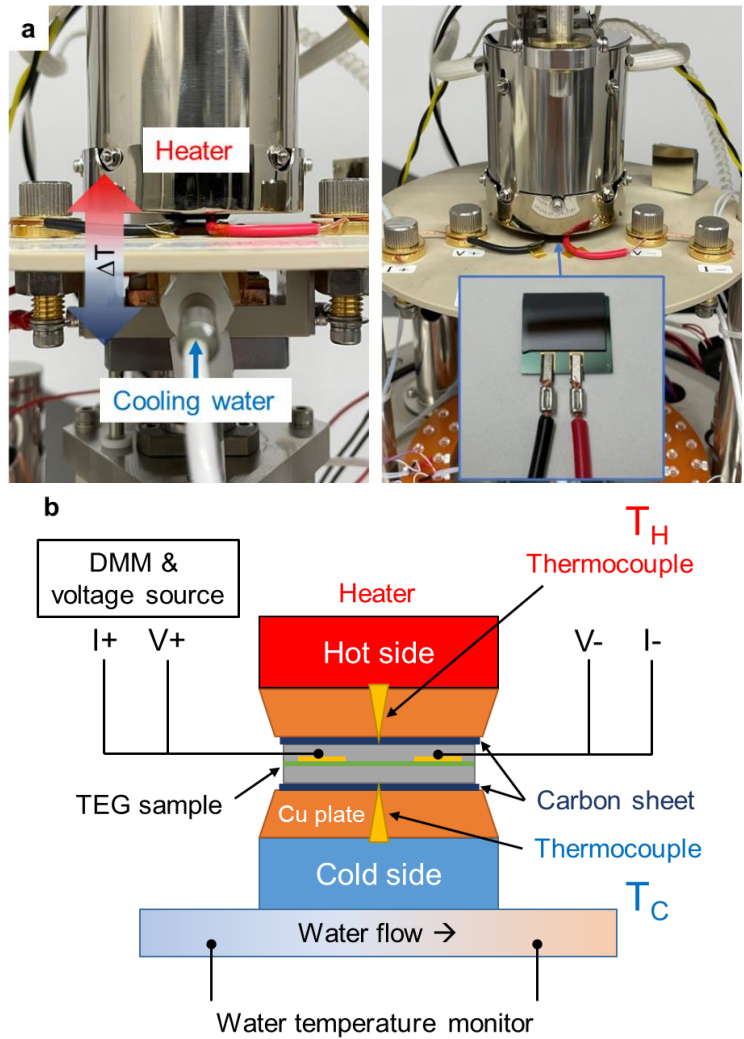

**Supplementary Figure 6 | TEG measurement setup.** (**a**) Photos of the measurement setup for TEG power composed of the cold stage, heater, and electrical connections. The inset shows a planar-type Si TEG with two wires. (**b**) Schematic of the measurement system. Thermoelectric voltage and electrical resistance are measured while applying controlled ΔT.

p. 22

**6. Thermal design of energy harvesting module**

We designed and developed the module shown in Supplementary Fig. 7 for the in-field energy harvesting experiment. The thermal design of the module is important to efficiently apply a temperature difference $\Delta T_{TEG}$ to the TEG device from the temperature difference $\Delta T_{ambient}$ in the environment to obtain a large amount of power generation. The thermal resistance of the TEG chip is estimated as shown in Supplementary Table 3 when the TEG chip is considered an effective material. The dimensions of the case body were designed to accommodate the energy harvesting circuits, sensors, and wireless communication modules while minimizing the thermal resistance of the Al blocks. A thermally conductive interface material (TIM) was used between the Al block and the TEG to reduce solid-gas interfaces due to surface roughness so that the thermal resistance decreases. The ratio of $\Delta T_{TEG}/\Delta T_{ambient}$ is expressed as the ratio of the thermal resistance of the TEG to the whole series thermal resistance, which is 50 %. This value was used in the analysis of in-field energy harvesting experiments.

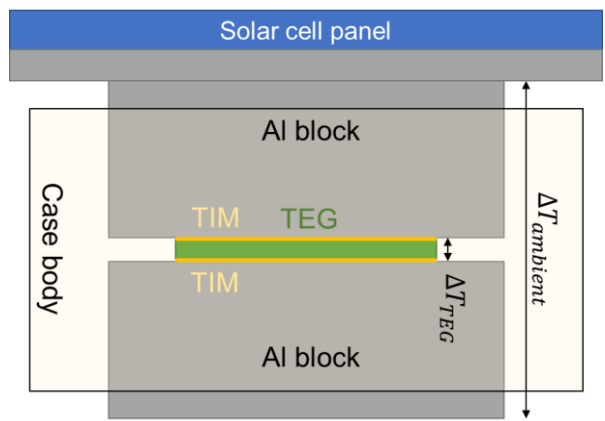

**Supplementary Figure 7 | Schematic of energy harvesting module.** Cross-section image of the schematic energy harvesting module on the solar cell panel. The module is composed of the Si TEG, two Al blocks, and two thermal interfacial material (TIM) layers.

**Supplementary Table 3 | Material properties used in energy harvesting module.**

|  | $\kappa$ (Wm$^{-1}$K$^{-1}$) | height (mm) | area (cm$^2$) | Thermal resistance (KW$^{-1}$) |
|---|---|---|---|---|
| TEG | 10 (effective) | 1 | 10 | 0.1 |
| Al block | 240 | 21 | 25 | 0.035 |
| TIM | 1.5 | 0.02 | 10 | 0.0133 |

p. 23